\documentclass[seceq]{ptptex}

\usepackage{graphicx}

\title{
Cosmological Constraints for the Cold Dark Matter and Model Building based on the Flavor Symmetric Radiative Seesaw Model\footnote{
Talk at ICGA8-For The 100th Anniversary of Hideki Yukawa and Promotion of Women Scientists-, Nara Women's University, Japan (August 29-September 1, 2007).
The title is changed.}
}

\author{
Hiroshi \textsc{Okada\footnote{I would like to thank J.~Kubo and Y.~Kajiyama, who are my collaborators, for useful discussions.}}
}

\inst{
Institute for Theoretical Physics,
Kanazawa University,
Kanazawa 920-1192, Japan
okada@hep.s.kanazawa-u.ac.jp
}

\abst{
 It is now clear that the masses of the neutrino sector are much lighter than those of the other three sectors.There are many attempts to explain the neutrino masses radiatively by means of inert Higgses, which don't have vacuum expectation values. Then one can discuss cold dark matter candidates, because of no needing so heavy particles and having a $Z_2$ parity symmetry corresponding to the R-parity symmetry of the MSSM.  
   The most famous work would be the Zee model\cite{zee}.  
  
 Recently a new type model\cite{ma1}  along this line of thought was proposed by Mr. E. Ma.
We paid attention to this idea. We introduce a flavor symmetry based on a dihedral group $D_6$ \cite{chris} to constrain the Yukawa sector. For the neutrino sector, we find that the maximal mixing of atmospheric neutrinos is realized, it can also be shown that only an inverted mass spectrum, the value of $|V_{MNS_{13}}|$ is 0.0034 and so on. When one extends the Higgs sector, it leads to FCNCs mediated by Higgs fields generally. But in our model,
the FCNCs are (of course) suppressed for the experiments sufficiently.\cite{mondra}   
For the fermionic CDM candidates, we find that the mass of the CDM and the inert Higgs should be larger than about $230$ and $300$ GeV, respectively.
If we restrict ourselves to a perturbative regime, they should be lighter than about $750$ GeV.\cite{oka1} 
}

\begin{document}

\maketitle

\section{Model building}

 Fermionic and bosonic fields are assigned as Table1 and Table2 respectively. 

\begin{table}[thb]
\begin{center}
\begin{tabular}{|c|cccccc|} \hline
 & $L_S$ & $n_S$ & $e^c_S $&$L_I$&$n_I$&$e^c_I$ 
  \\ \hline
 $SU(2)_L\times U(1)_Y$ 
 & $({\bf 2}, -1/2)$  &  $({\bf 1}, 0)$  &  $({\bf 1}, 1)$
 & $({\bf 2}, -1/2)$&  $({\bf 1}, 0)$
 &  $({\bf 1}, 1)$
  \\ \hline
 $D_6$ & ${\bf 1}$  &  ${\bf 1}'''$  &  ${\bf 1}$
 & ${\bf 2}'$&  ${\bf 2}'$&  ${\bf 2}'$
 \\ \hline
 $\hat{Z}_2$
 & $+$ &$+$  & $-$  & $+$ 
 &$+$  & $-$
  \\ \hline
   $Z_2$
 & $+$ &$-$  & $+$  & $+$ &$-$  & $+$ 
   \\ \hline
\end{tabular}
\caption{The $D_6 \times \hat{Z}_2\times Z_2$ 
assignment for the leptons.  The subscript $S$ indicates
a $D_6$ singlet, and the subscript $I$ running from $1$ to $2$
stands for a $D_6$ doublet. $L$'s denote the $SU(2)_L$-doublet leptons,
while $e^c$ and $n$ are the $SU(2)_L$-singlet leptons.
}
\end{center}
\end{table}
\begin{table}[thb]
\begin{center}
\begin{tabular}{|c|cccc|} \hline
 & $\phi_S$ &$\phi_I$ & $\eta_S$&$\eta_I $
  \\ \hline
   $SU(2)_L\times U(1)_Y$ 
 & $({\bf 2}, -1/2)$  &  $({\bf 2}, -1/2)$   &  $({\bf 2}, -1/2)$ 
 & $({\bf 2}, -1/2)$ 
  \\ \hline
 $D_6$ & ${\bf 1}$ &${\bf 2}'$  &  ${\bf 1}'''$  & ${\bf 2}'$
 \\ \hline
 $\hat{Z}_2$ &$+$ 
 & $-$ &$+$  & $+$ 
  \\ \hline
   $Z_2$
 & $+$ &$+$  & $-$  & $-$ 
   \\ \hline
\end{tabular}
\caption{The $D_6 \times \hat{Z}_2\times Z_2$ 
assignment  for the $SU(2)_L$  Higgs doublets.
}
\end{center}
\end{table}
Under  $Z_2$  (which plays the role of $R$ parity
in the MSSM), only the right-handed neutrinos $n_S, n_I$ and 
the extra Higgs $\eta_S, \eta_I$ are odd.
The quarks are assumed to belong to ${\bf 1}$ of $D_6$ 
with $(+,+)$ of  $\hat{Z}_2\times Z_2$ so that the 
quark sector is basically the same as the SM, where 
the $D_6$ singlet Higgs $\phi_S$ with
$(+,+)$ of  $\hat{Z}_2\times Z_2$ plays the role of the SM Higgs
in this sector. No other Higgs can couple to the quark sector at the tree-level.
In this way we can avoid tree-level FCNCs in the quark sector.
So, $\hat{Z}_2$ is introduced to forbid  tree-level couplings of 
the $D_6$ singlet Higgs $\phi_S$ with the leptons
and simultaneously to forbid  tree-level couplings of $\phi_I, \eta_I$ and $\eta_S$
with the quarks.

\section{Lepton masses and mixing}

The most general renormalizable $D_6 \times \hat{Z}_2 \times 
Z_2$ invariant 
Yukawa interactions in the leptonic sector 
can be gained. By the Higgs mechanism, the charged lepton and the neutrino masses are generated from the $S_{2}$
invariant VEVs \cite{oka2}, and the mass matrix becomes
\begin{eqnarray}
{\bf M}_{e} = \left( \begin{array}{ccc}
-m_{2} & m_{2} & m_{5} 
\\  m_{2} & m_{2} &m_{5}
  \\ m_{4} & m_{4}&  0
\end{array}\right)
\label{mlepton}
,
{\bf M}_{\nu} = \left( \begin{array}{ccc}
2 (\rho_{2})^2 & 0 & 
0
\\ 0 & 2 (\rho_{2})^2 & 2 \rho_2 \rho_{4}
  \\ 0 & 2 \rho_2 \rho_{4}  &  
2 (\rho_{4})^2 +
(\rho_3)^2\exp i 2 \varphi_{3}
\end{array}\right),
\label{m-nu}
\nonumber
\\
\end{eqnarray}
where all the mass parameters
appearing in (\ref{mlepton}) can be assumed to be real.

Now we can lead some predictions for the lepton secotor.
\begin{itemize}
\item First, since the mixing of atmospheric neutrinos must be maximal form from the experiments, and only an inverted mass spectrum can be allowed.   
\item Second,\[U_{e3}\sim0.0034<<0.2  .\]
\item Third and Fourth, 
 \[ m_{\nu2,min.}\sim f(\tan\theta_{sol.},\Delta m^2_{32},\Delta m^2_{12},\phi=0)=0.038\sim0.067 eV,\]
\[ m_{ee,min.}\sim g(\tan\theta_{sol.},\Delta m^2_{32},\Delta m^2_{12},\phi=0)=0.034\sim0.069 eV.\]
\end{itemize}

where $U_{e3},m_{\nu2,min.},m_{ee,min.},\theta_{sol.},\Delta m_{32},\Delta m_{12},$
and $\phi$ mean the Maki  Nakawgawa matrix, the minimal second neutrino mass, the minimal effective majorana mass, the solar mixing angle, the atmospheric mass difference, the solar mass difference and a phase respectively.

\section{Cold Dark Matter } 
  I will move on to the discussion of the CDM .
  Where I will suppose the CDM , which is fermionic. 
Based on our model, we can consider $\mu$ $\rightarrow $ e, $\gamma$ diagram mediated only by the charged extra Higgs eta exchange. As a result of the calculation,
I find that it is more natural that ns remains as a fermionic CDM candidate.
Otherwise I have to impose a fine tuning for $n_I$ mass to sufficiently suppress the $\mu$ $\rightarrow $ e, $\gamma$ process. Furthermore I found that almostly charged extra Higgs $\eta_S$ couples to $e_L$ and $n_S$ owing to our original matrix. Therefore  there would be a clean signal if the charged extra Higgs $\eta_S$ was produced at LHC !

In the next, we would like to investigate whether or not $n_S$ can be a good CDM candidate from the cosmology. 
We found that the $n_S$ is annihilated mostly into an $e^+-e^-$ pair and a $\overline{\nu}_{\tau}-\nu_{\tau}$ pair in this model.
Reffering the following papers \cite{griest1}\cite{griest2}, we can compute the relativistic cross section.

\begin{figure}[htb]
\includegraphics*[width=0.5\textwidth]{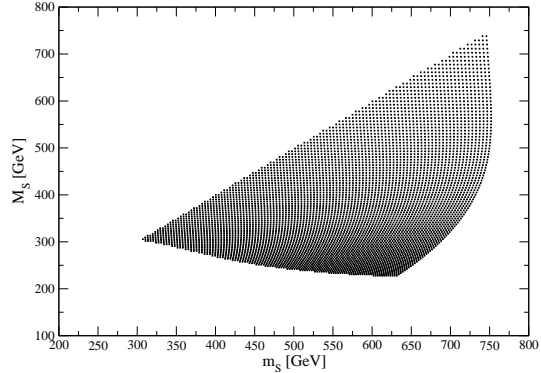}
\caption{\label{ms-ms}\footnotesize
The region in the $m_S-M_S$ plane
in which $\Omega_d h^2=0.12,
B(\mu\to e \gamma) < 1.2\times 10^{-11}$ and
$| h_3| <1.5$ are satisfied.
}
\end{figure}

 In fig. \ref{ms-ms} we present
the allowed region in the $m_S-M_S$ plane, in which
$\Omega_d h^2=0.12$
and $B(\mu\to e \gamma) < 1.2\times 10^{-11}$
are satisfied, where we assume $|h_3| <1.5$
If we allow larger $|h_3|$,
then the region expands  to
 larger $m_S$ and $M_S$,
 and for  $|h_3|\sim 0.8$ there is no allowed region.
 As we can also see from fig. \ref{ms-ms},
the mass of the CDM and the mass of the inert Higgs should be larger than 
about $230$ and $300$ GeV, respectively.
If we restrict ourselves to
a perturbative regime, they should be lighter than about $750$ GeV.

In the last analysis, we calculated the mass bound for Sunyaev-Zel'dovich(SZ) effect.\cite{s-z} In our model, $\eta^+_s$, which decays to $e_L$ that has high energy, may affects the CMB by the Compton scattering, if the life time isn't between $10^{-5}$-$10^{-7}$s. The condition that $\eta^+_s$ comes into the allowed life time region, mass($m_S$) can be given by 
\[ 30GeV<m_S<750GeV,\]
where the Yukawa coupling nearly equals to 1, and $m_S\sim m_{e_L}>> M_S$ are assumed. As a result of the analysis, I find that the SZ effect satisfy the both constraints of $\mu\to e\gamma$ and cosmological pair annihilation for CDMs sufficiently.

\section{Conclusions} 
We can conclude that; 
\begin{itemize}
\item we could construct the predictive model for the neutrino sector radiatively.
\item from the $\mu \to e \gamma$, cosmological pair annihilation for CDMs and SZ effect, if CDMs are fermionic, and we could single out the $D_6$ sym. singlet right-handed neutrino($n_S$) as the best CDM candidate.  
\item an inert Higgs with a mass between 300 GeV and 750 GeV decays mostly intoan electron (or positron) with a large missing energy, where the missing energy is carried out by the CDM candidate.
\item $m_S$ bound is satisfied with the restrictions($m_S=30\sim 750GeV$) coming from the Z effect. 
\item this dominant mode($\eta^+_s\to n_S,e_L$)  would be a clean signal at LHC.  
\end{itemize}
In our further discussion, $\eta^+_s$ may be able to become the solvable origin for the Lithium problem.\cite{kusa} 

\section*{Acknowledgements}
I would like to thank organizers, especially Prof. Kenmoku, in Nara Women's Univ,
and Dr. M. Kusakabe for useful discussions.
This work is supported by Nara Women's University (NWU),
Japan Society for the Promotion of Science (JSPS),
Asia Pacific Center for Theoretical Physics (APCTP),
Inoue Foundation for Science,
Epson International Scholarship Foundation,
Physical Society of Japan (JPS),
The Astronomical Society of Japan (ASJ),
Yukawa Institute for Theoretical Physics, Kyoto University (YITP) and
Nara Convention Bureau (NCB)
%

\end{document}